# Pressure tuning of light-induced superconductivity in $K_3C_{60}$

A. Cantaluppi[1,2,♣], M. Buzzi[1,♣], G. Jotzu[1], D. Nicoletti[1,2], M. Mitrano[1], D. Pontiroli[3], M. Riccò[3], A. Perucchi[4], P. Di Pietro[4], A. Cavalleri*[1,2,5]

[1]Max Planck Institute for the Structure and Dynamics of Matter, Hamburg, Germany

[2]The Hamburg Centre for Ultrafast Imaging, Hamburg, Germany

[3]Dipartimento di Scienze Matematiche, Fisiche e Informatiche, Università degli Studi di Parma, Parma, Italy

[4]INSTM UdR Trieste-ST and Elettra–Sincrotrone Trieste, Trieste, Italy.

[5]Department of Physics, Oxford University, Clarendon Laboratory, Oxford UK.

*email: andrea.cavalleri@mpsd.mpg.de

**Optical excitation at terahertz frequencies has emerged as an effective means to manipulate complex solids dynamically. In the molecular solid $K_3C_{60}$, coherent excitation of intramolecular vibrations was shown to transform the high-temperature metal into a non-equilibrium state with the optical conductivity of a superconductor. Here we tune this effect with hydrostatic pressure, and we find it to gradually disappear around 0.3 GPa. Reduction with pressure underscores the similarity with the equilibrium superconducting phase of $K_3C_{60}$, in which a larger electronic bandwidth is detrimental for pairing. Crucially, our observation excludes alternative interpretations based on a high-mobility metallic phase. The pressure dependence also suggests that transient, incipient superconductivity occurs far above the 150 K hypothesised previously, and rather extends all the way to room temperature.**

♣ These authors contributed equally to this work.

Resonant optical excitation of infrared-active phonon modes can drive the crystal lattice of solids nonlinearly[1], excite other orders coherently[2], switch lattice polarization[3], drive insulator-to-metal[4] or magnetic transitions[5], and even induce transient superconductivity above equilibrium $T_c$[6,7]. In the potassium-doped fulleride $K_3C_{60}$, a superconductor at temperatures below $T_c$ = 20 K, excitation of local molecular vibrations was shown to induce superconducting-like optical properties in the high temperature metal (T > $T_c$)[8]. Key features of this state are a ~12-meV-wide gap in the resistive optical conductivity $\sigma_1(\omega)$, twice as large as the equilibrium 6-meV-wide superconducting gap, and a divergent low-frequency imaginary conductivity $\sigma_2(\omega)$, indicative of large carrier mobility. This state was found to extend to at least 100 K, hence up to temperature scales far in excess of equilibrium $T_c$ (20 K). For T > 100 K, a partially gapped state was reported. In this higher temperature regime, the light-induced state can be interpreted either as a high mobility metal "en route" to a transient superconducting state, or as an incipient superconductor, which is only partially coherent.

Many theoretical mechanisms have been invoked to explain these observations, ranging from a dynamical reduction of the electronic bandwidth[9], to the parametric amplification of the pairing instability[10] and to electron attraction[11] in vibrationally excited molecular sites[12,13]. Recent experiments in bi-layer graphene are consistent with some of these suggestions[9,10], as the optical excitation of a similar vibrational mode as that driven in $K_3C_{60}$ appears to increase the electron phonon interaction[14]. Finally, recent theoretical work has raised the possibility that photo-stimulation may involve optical excitation and cooling of above-gap thermal quasi-particles into a super-excitonic state with high electronic heat capacity[15].

The face-centred cubic structure of doped fullerides $A_3C_{60}$ is shown in Fig. 1a. Three electrons are donated by the alkali atoms to each $C_{60}$ molecule, which then form three narrow, half-filled bands near the Fermi level. Figure 1c shows how the equilibrium superconducting transition appears in the steady-state optical properties of $K_3C_{60}$, measured above and below the critical temperature $T_c$

= 20 K. When cooling metallic $K_3C_{60}$ (red curves) below $T_c$, one observes large changes in the optical properties: a saturation of the low-frequency reflectivity to R = 1, a 6 meV gap in the real part of the optical conductivity $\sigma_1(\omega)$, and a $1/\omega$ divergence in the imaginary part $\sigma_2(\omega)$[16,8] (blue curves).

As superconductivity in $K_3C_{60}$ emerges from a combination of Jahn-Teller intramolecular distortions and electronic correlations[17,18,19], it is natural to explore the response of the material to direct excitation of optically accessible vibrational modes. In Fig. 1d we report the optical properties of polycrystalline powders of $K_3C_{60}$, 1 ps after the excitation tuned to "on ball" infrared active modes of $T_{1u}$ symmetry at 170 meV energy (7.3 μm wavelength), whose atomic distortion is displayed in Fig. 1b. At this frequency, strongly correlated metallic carriers are also excited, as the radiation is also resonant with a broad absorption peak extending from ~40 to 200 meV[20], whose precise origin is still unclear[15]. A broadband probe pulse was used to detect the light-induced changes in the optical reflectivity and complex optical conductivity between 1.6 and 7 THz (6.5 – 29 meV) using THz-time-domain-spectroscopy. Starting from the unperturbed metallic state at 100 K (red curves), we observed an increase in the reflectivity, which saturates to R = 1 for all probe photon energies below ~12 meV, a gapped $\sigma_1(\omega)$ and a divergent $\sigma_2(\omega)$. These data confirm the results of Ref. 8, but were recorded with an improved apparatus, involving higher pump fluence and a broader probe bandwidth (see Supplementary Information S3).

In this paper, we study how the features reported in Fig. 1d change with the application of hydrostatic pressure. At equilibrium, the application of pressure reduces the superconducting transition temperature $T_c$, because of the increase in the electronic bandwidth when the intermolecular spacing is reduced[21,22]. As shown in Fig. 2, the size of the optical gap ($2\Delta_0$) and the critical temperature[23,24] decay linearly already at relatively modest pressures. Due to the low bulk modulus (28 GPa[24]), a pressure of 3 GPa reduces the superconducting gap to less than half of the ambient-pressure value, as the electronic bandwidth increases by ~25%[25].

Figure 3 displays snapshots of the measured optical reflectivity R(ω) at the sample-diamond interface, along with complex conductivity spectra, $\sigma_1(\omega)$ and $\sigma_2(\omega)$, for different values of static pressure. The exact pressure was measured with calibrated ruby fluorescence (see Supplementary Information S3). In each panel, the red and blue curves trace the optical properties of the equilibrium metal and those of the non-equilibrium state induced by photo-excitation, respectively. For pressures up to 0.17 GPa (Fig. 3a-c) the transient optical response of $K_3C_{60}$ is similar to that observed at ambient pressure, with a reflectivity approaching R=1, a gapped $\sigma_1(\omega)$ and a divergent $\sigma_2(\omega)$ toward low frequencies. However, some spectral weight is also found in $\sigma_1(\omega)$ at low energies, indicative of reduced coherence.

As the applied pressure increases, a stronger suppression of the light-induced changes in both the reflectivity and complex optical conductivity is observed (Fig. 3d-e). Above 0.3 GPa the enhancement in the reflectivity is clearly less pronounced, and a progressively broader Drude peak appears at low frequency in the $\sigma_1(\omega)$ spectrum.

The reduction in light-induced coherence observed as a function of pressure is clearly not compatible with what expected for a light-induced metallic state, as a lattice compression in a metal is typically associated with larger electronic bandwidth, smaller effective mass and larger mobility. This is for example evident when analysing the equilibrium metallic properties in the red curves of Fig. 3 (see also Supplementary Information S2), where one observes larger plasma frequencies $\omega_p$ with increasing pressure.

In Fig. 4a-c we report the fractional spectral weight loss for frequencies inside the gapped region of the spectrum, obtained by integrating $\sigma_1(\omega)$ between 6.5 and 12.9 meV for different pressures and base temperatures of 100 K, 200 K and 300 K (see Supplementary Information S9 for full data sets at 200 and 300 K). Shaded blue areas indicate the pressure-temperature ranges where the light-induced state is gapped. Overall, the light-induced gap fills already at moderate pressure values, becoming even smaller for increasing temperature. For $P \gtrsim 0.3$ GPa, the pressure dependence of the light-induced effects is strongly reduced.

Fits to the optical properties of Fig. 3 make the qualitative analysis above quantitatively significant (see Supplementary Information S10). By fitting the transient optical response of $K_3C_{60}$, we extrapolated the value of the low frequency optical conductivity $\sigma_0 = \lim_{\omega \to 0} \sigma_1(\omega)$. To compare both superconducting-like and metallic-like states in a consistent fashion, we used a Drude-Lorentz fit for the entire regime, in which $\sigma_0$ was allowed to float from finite (metal) to infinite values (perfect conductor), and a single lorentzian was used to capture the mid-infrared absorption band extending from 40 to 200 meV.

The results of this analysis are summarised in Fig. 5a-c, where we report the pressure dependence of $\sigma_0$ for three temperatures (100 K, 200 K, and 300 K). The red squares describe the pressure dependence for the optical properties of the equilibrium metal, while the blue diamonds that of the photoexcited state. As shown in these plots, the equilibrium metallic conductivity increases with applied pressure. On the contrary, two pressure regimes are found for the light-induced state, one in which $\sigma_0$ *decreases* for small pressures (d$\sigma_0$/d$P$ < 0, blue shaded area) and one where it eventually *increases* slightly for higher pressures (d$\sigma_0$/d$P$ > 0, yellow shaded area). Several indications can be extracted from these data.

First, as mentioned above, from the optical properties alone reported in Ref. 8, one could not uniquely differentiate a superconductor from a perfect conductor, as optics only identifies the density of charge carriers and the scattering rate. The hydrostatic pressure dependence reported here adds crucial information. At low pressures, the photoexcited state has clear superconducting-like pressure dependence (d$\sigma_0$/d$P$ < 0), whereas for higher pressures the response is clearly metal-like (d$\sigma_0$/d$P$ > 0). Furthermore, at the high pressures $\sigma_0$ of the photo-excited state follows the same slope as that of the equilibrium metal.

In this context the results reported for high temperatures (T = 200 K and T = 300 K) are surprising. In that temperature range, a high mobility metallic state was proposed to interpret the data of Ref. 8. However, this interpretation was also not unique, as a superconducting-like state with progressively lower coherence could also have explained the data. Figure 5 suggests that in the low pressure regime the d$\sigma_0$/d$P$ < 0 behaviour is retained all the way to 300 K, hinting that some incipient features of transient superconductivity may already be present up to room temperature.

These observations also provide guidance for a microscopic explanation of our results. Indeed, as summarised in Fig. 6, one finds a very strong dependence of the light-induced optical conductivity on pressure, and for the higher-pressure ranges (smaller lattice constants) the metallic phase (yellow) is stabilised. This is compatible with the interpretations reported in Refs. 9,10,11,27,28. It is less clear if the quasi-particle cooling scenario[15] would exhibit such a dramatic dependence on the electronic bandwidth, as the equilibrium mid-infrared absorption band appears to be only marginally affected by pressure. Figure 6 also indicates a clear path for future research in the broader context of $A_3C_{60}$ superconductivity, showing on the right hand side the region of the phase diagram still to be accessed (panel b), with the interesting perspective of optimizing light-induced superconductivity further for even larger lattice spacing.

**Acknowledgments**

The research leading to these results received funding from the European Research Council under the European Union's Seventh Framework Programme (FP7/2007-2013)/ERC Grant Agreement no. 319286 (QMAC). We acknowledge support from the Deutsche Forschungsgemeinschaft via the excellence cluster "The Hamburg Centre for Ultrafast Imaging - Structure, Dynamics and Control of Matter at the Atomic Scale" and the priority program SFB925. M. Buzzi acknowledges financial support from the Swiss National Science Foundation through an Early Postdoc Mobility Grant (P2BSP2_165352).

**REFERENCES (Main Text)**

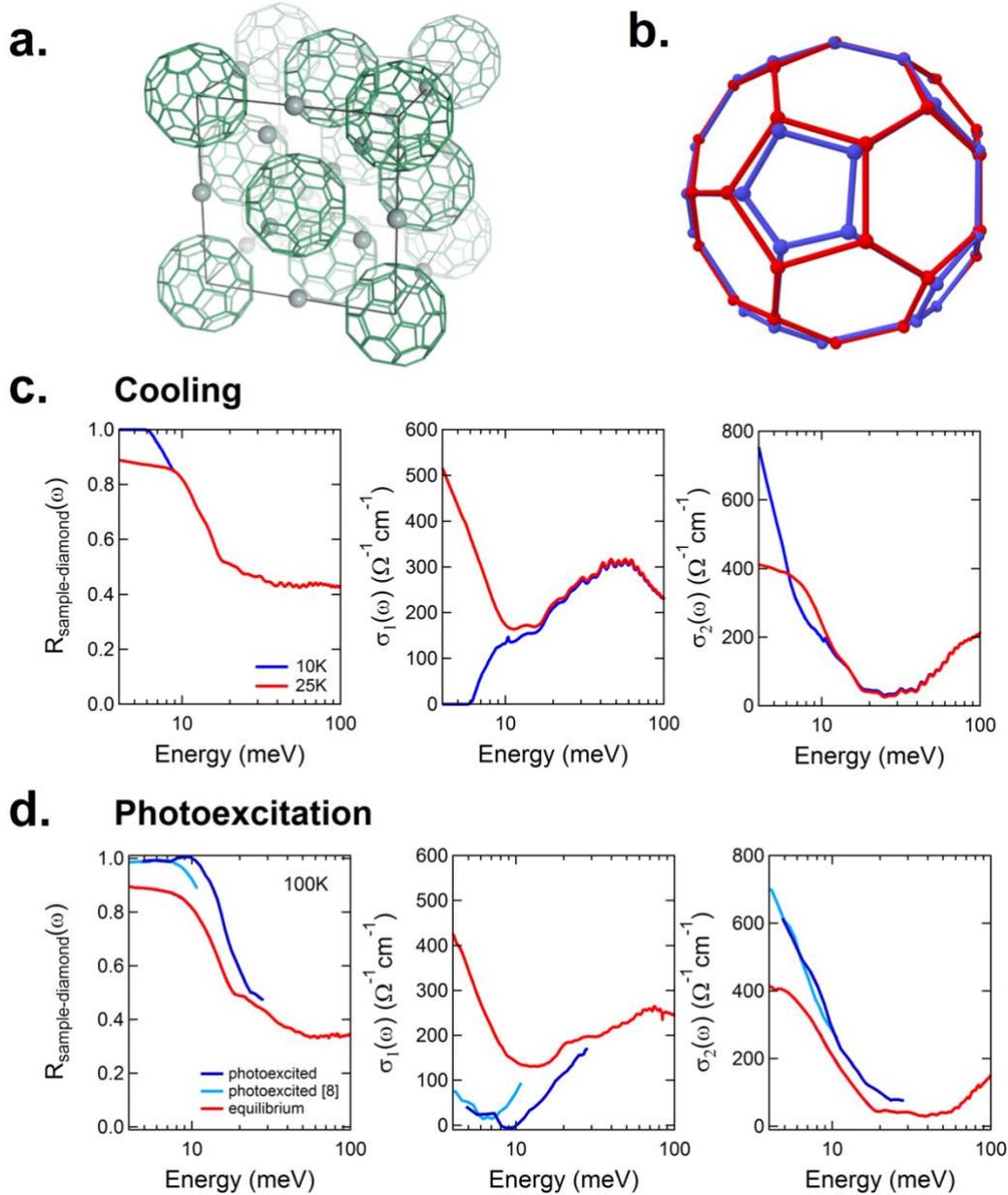

**Figure 1. Structure, equilibrium phase transition, and transient light-induced phase of $K_3C_{60}$. a.** The f.c.c. crystal structure of $K_3C_{60}$ (Ref. 29). The $C_{60}$ molecules are represented by the green bonds connecting all the C atoms. The grey spheres are the K atoms, acting as spacers between neighbouring buckyballs. The equilibrium lattice constant is 14.26 Å at room temperature. **b.** $C_{60}$ molecular distortion (blue) along the $T_{1u}(4)$ vibrational mode coordinates. The equilibrium structure is shown in red. **c.** Reflectivity (sample-diamond interface), real and imaginary optical conductivity - $R(\omega)$, $\sigma_1(\omega)$ and $\sigma_2(\omega)$ - across the equilibrium superconducting transition in $K_3C_{60}$. The red curves are measured at 25 K, in the metallic phase. The blue curves refer to the equilibrium superconductor (10 K). **d.** Reflectivity (sample-diamond interface), real and imaginary optical conductivity - $R(\omega)$, $\sigma_1(\omega)$ and $\sigma_2(\omega)$ - of $K_3C_{60}$ at equilibrium (red) and 1 ps after photo-excitation (blue) at T = 100 K. The light-blue curves show the data reported in Ref. 8, measured with a fluence of 1 mJ/cm$^2$, while those in dark blue were measured with a broader probe spectrum and a higher pump fluence (3 mJ/cm$^2$).

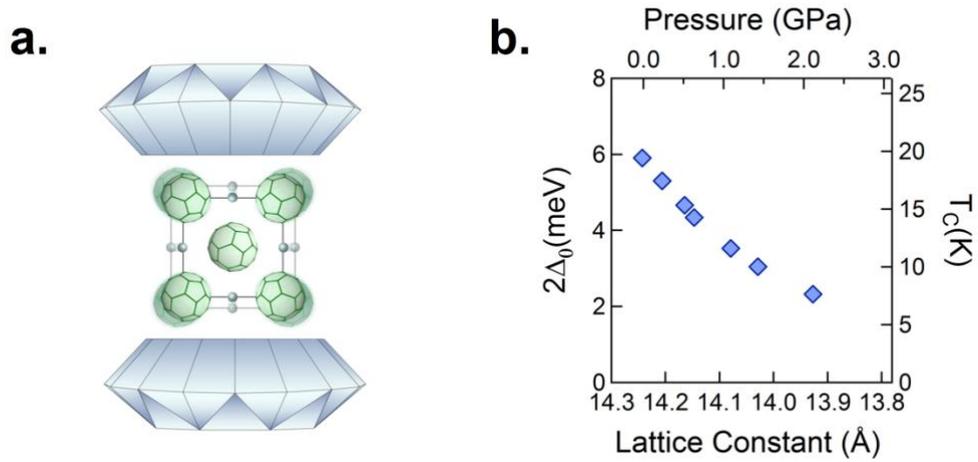

**Figure 2. Equilibrium pressure dependence of superconducting K$_3$C$_{60}$. a.** Schematic representation of a diamond anvil cell (DAC) acting on the K$_3$C$_{60}$ crystal structure. By applying external pressure, the inter-molecular distances get reduced. **b.** Superconducting transition temperature and calculated optical gap ($2\Delta_0/k_BT = 3.52$[23]), plotted as a function of lattice parameter and external pressure. Data adapted from Refs. 24, 30.

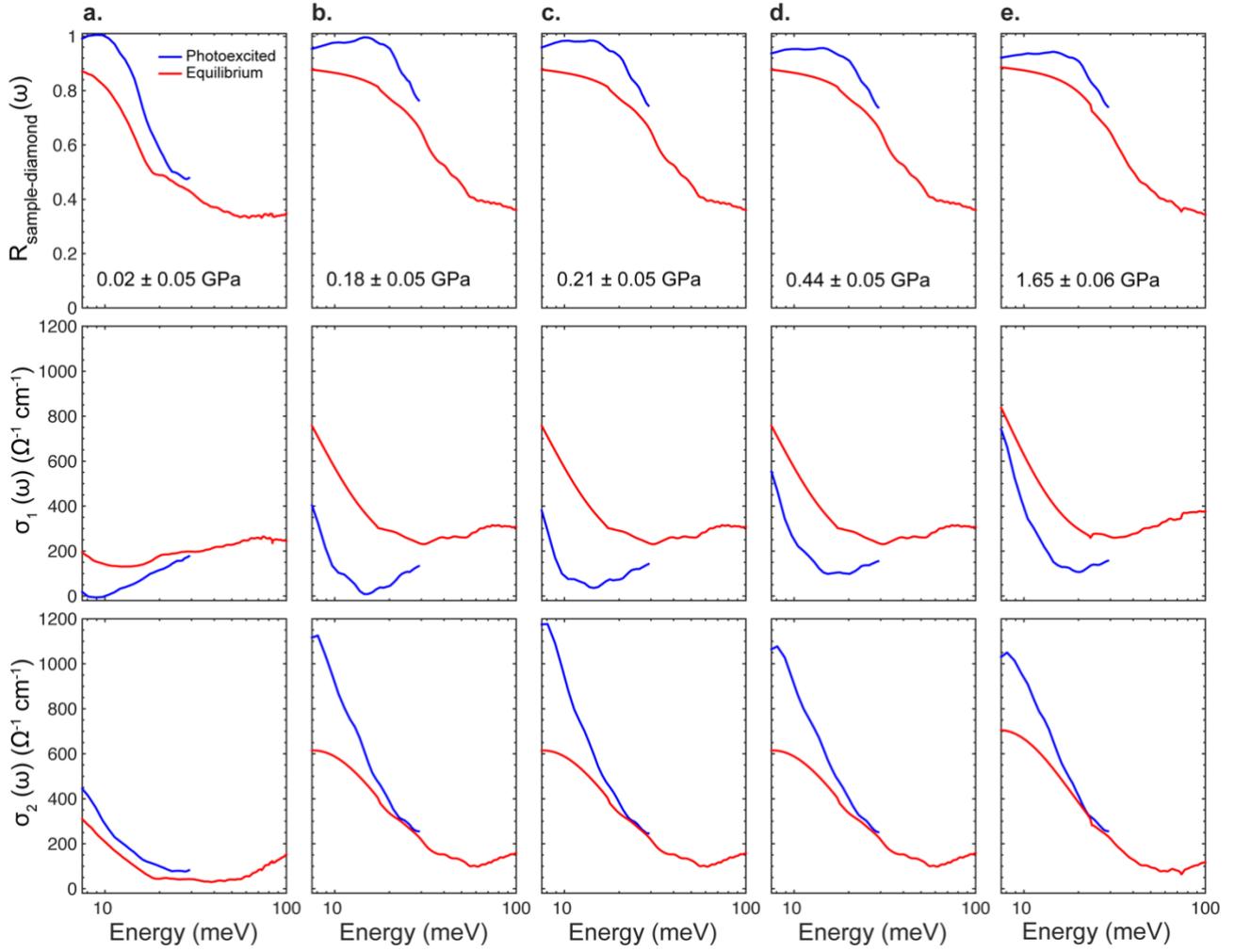

**Figure 3. Pressure dependence of the transient optical properties of $K_3C_{60}$ at T = 100 K.** Reflectivity (sample-diamond interface) and complex optical conductivity of $K_3C_{60}$ measured at equilibrium (red) and 1 ps after photoexcitation (blue) at T = 100 K, for different external hydrostatic pressures. All data were taken with the same pump fluence of 3 mJ/cm$^2$.

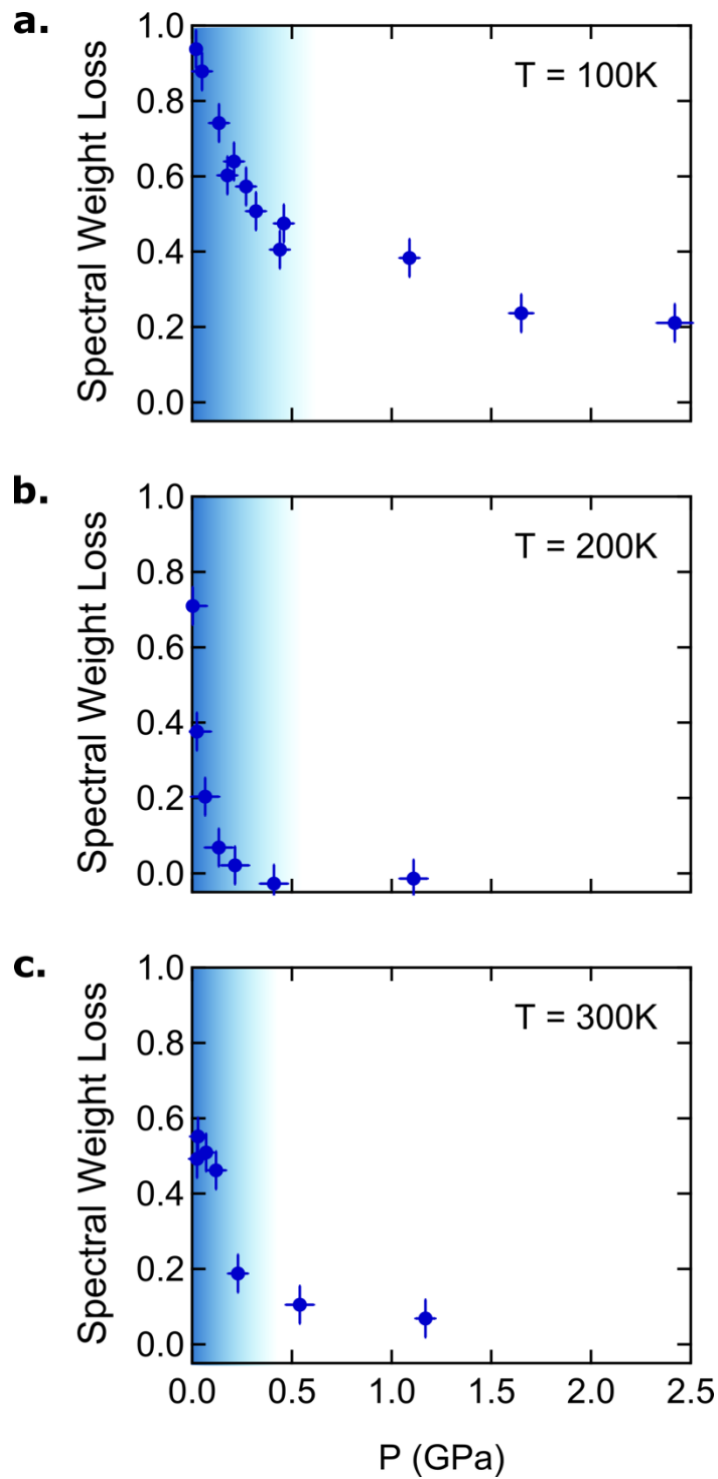

**Figure 4. Scaling of the $\sigma_1(\omega)$ gap with external pressure.** Photo-induced reduction in $\sigma_1(\omega)$ spectral weight, integrated between 6.5 and 12.9 meV, normalised by the equilibrium value (integrated in the same range). The blue shaded areas identify the regions in which a photo-induced conductivity gapping is measured. All data were taken with the same pump fluence (3 mJ/cm$^2$). Vertical error bars represent uncertainties determined from different sets of measurements and horizontal error bars show the calibration uncertainty of the ruby fluorescence measurements used to determine the pressure (see Supplementary Information S3).

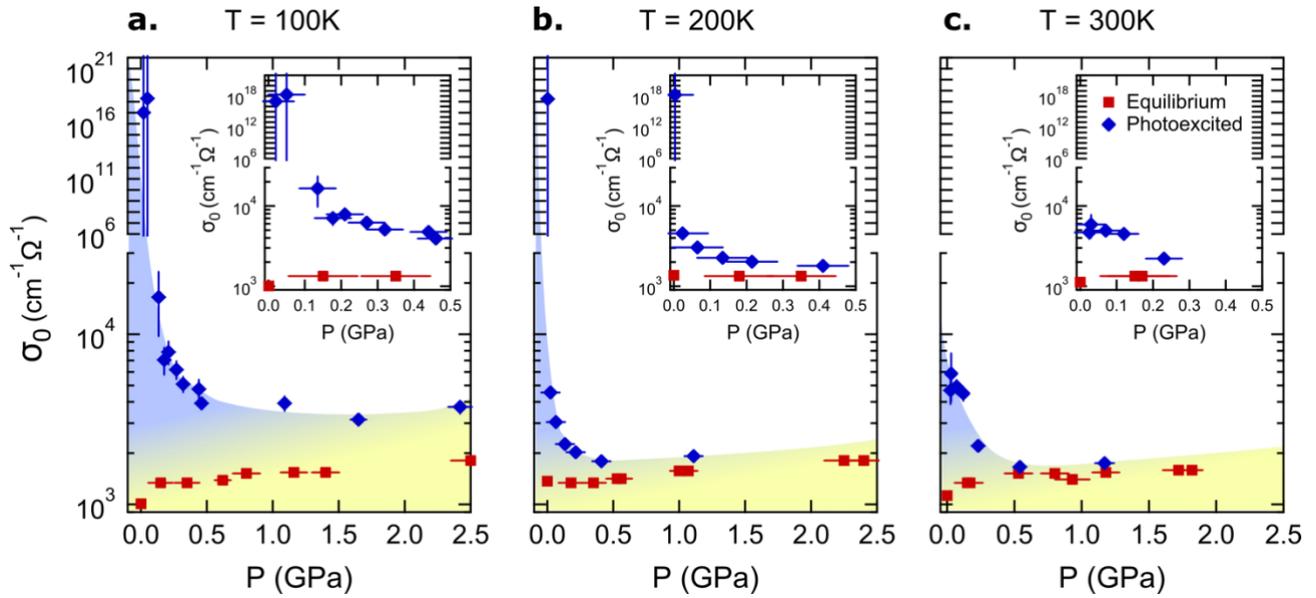

**Figure 5. Pressure dependence of the extrapolated conductivity.** Blue diamonds are extrapolated zero-frequency conductivities extracted from Drude-Lorentz fits of the transient optical spectra, as a function of pressure and for three different temperatures: 100 K (**a**), 200 K (**b**), and 300 K (**c**). Red squares are the corresponding zero-frequency conductivities determined at equilibrium. The blue areas identify the regions in which $\sigma_0$ is suppressed by pressure ($d\sigma_0/dP < 0$) while those in yellow indicate the regime in which $d\sigma_0/dP > 0$. All data were taken with the same pump fluence (3 mJ/cm$^2$). The insets show a close-up of the low-pressure region. Vertical error bars reflect the fit uncertainty and horizontal error bars show the calibration uncertainty of the ruby fluorescence measurements used to determine the pressure (see Supplementary Information S3).

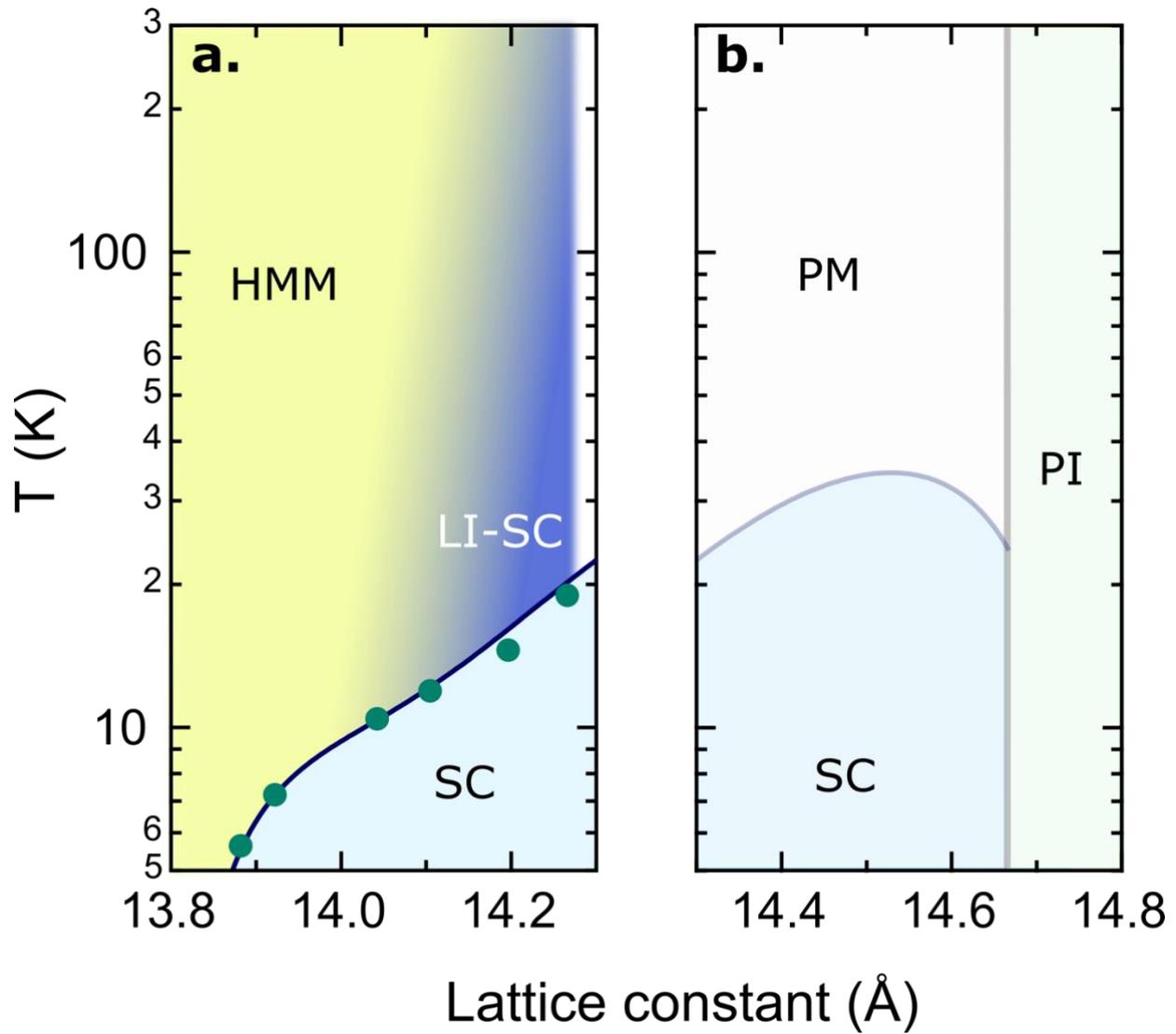

**Figure 6. Out-of-equilibrium phase diagram of f.c.c. $K_3C_{60}$. a.** $K_3C_{60}$ phase diagram (T *vs* room-temperature lattice constant and pressure). Green filled circles indicate the superconducting transition temperature ($T_c$) measured for different lattice parameters[24] at equilibrium. The light-blue filled area defines the equilibrium superconducting phase (**SC**). The dark blue shading stands for the Light-Induced Superconductor (**LI-SC**), which evolves into a High-Mobility Metal (**HMM**) under the application of pressure (red area). **b.** Extension of the f.c.c. $A_3C_{60}$ phase diagram for larger lattice constants. PI, PM, and SC refer to the equilibrium paramagnetic insulating, paramagnetic metallic, and superconducting phase, respectively. Based on Ref. 17,31,32.

# Supplementary Information

**S1. Sample growth and characterisation.**

Stoichiometric amounts of finely ground $C_{60}$ powder and potassium metal were sealed in a cylindrical vessel and closed in a Pyrex vial under vacuum (~$10^{-6}$ Torr). The potassium was kept separated from the fullerene powder during the thermal treatment, therefore only potassium vapors came in contact with $C_{60}$. The two reagents were heated at 523 K for 72 h and then at 623 K for 28 h. The vessel was then opened and the recovered black powder was reground and pelletized. Afterwards, the pellets were further annealed at 623 K for 5 days. All described operations were performed in inert atmosphere (vacuum or Ar glove box with <0.1 ppm $O_2$ and $H_2O$). The final product, $K_3C_{60}$, of an average grain size of 100-400 nm, was characterized by laboratory powder X-ray diffraction and SQUID magnetometry (Supplementary Fig. S1).

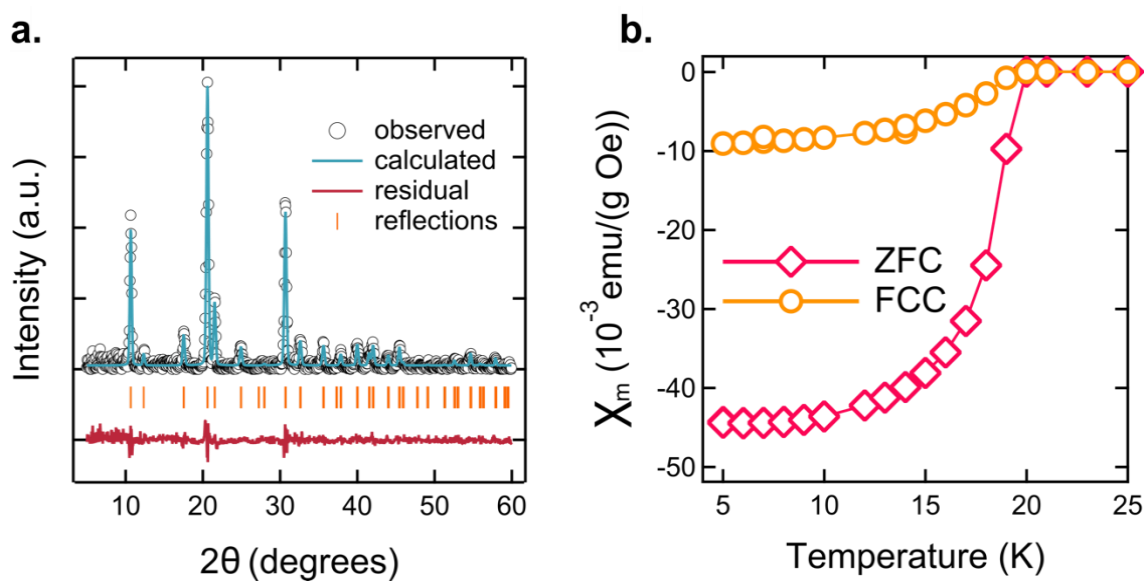

**Supplementary Figure S1. $K_3C_{60}$ sample characterisation**. (**a**) Powder X-ray diffraction data for the $K_3C_{60}$ used in this experiment together with a single f.c.c. phase Rietveld refinement. (**b**) Temperature dependence of the sample magnetic susceptibility with and without magnetic field (FCC: Field cooled cooling. ZFC: Zero field cooling).

**S2. Equilibrium optical response under pressure.**

The equilibrium optical properties of $K_3C_{60}$ were determined in a broad spectral range (5 - 500 meV) for different temperatures and pressures. Fourier-transform infrared spectroscopy measurement were carried out at the SISSI beamline (Elettra Synchrotron Facility, Trieste), using a commercial Bruker Vertex70 interferometer equipped with an Hyperion microscope[1]. Hydrostatic pressure was applied by a screw-driven opposing plate diamond anvil cell (DAC) using Boehler-Almax type IIac diamond anvils with a culet size of 2 mm. $K_3C_{60}$ powders (100 - 400 nm average grain size) were directly loaded in a 1 mm diameter sample compartment drilled in a 100 $\mu$m thick pre-indented Cu-gasket. This setup allowed achieving hydrostatic pressure values as high as 2.5 GPa, which were accurately measured *in situ* using a ruby fluorescence manometer[2,3]. The DAC was mounted on the cold finger of a liquid helium cryostat, which allowed varying the sample temperature. Note that all sample handling operations, including the loading of the $K_3C_{60}$ powders in the DAC, were performed in an argon atmosphere glove box with < 0.1 ppm $O_2$ and $H_2O$ in order to avoid oxygen contamination.

The equilibrium reflectivity of $K_3C_{60}$ was measured at the sample-diamond interface by referencing against a gold mirror placed into the holder at the sample position. The experimental curves were then extrapolated at low frequency ($\lesssim$ 5 meV) using Drude-Lorentz fits, while for high frequencies ($\gtrsim$ 500 meV) literature data on $K_3C_{60}$ single crystals[4] were used. We then performed a modified Kramers-Kronig transformation procedure for samples in contact with a transparent window[5], which allowed retrieving the complex optical conductivity of $K_3C_{60}$ at all temperatures and pressures.

The response measured at ambient pressure was already reported in Ref. 6, where differences and analogies with data taken on single crystals[4] were also discussed. Here, in Supplementary

Fig. S2, we show the equilibrium optical properties of $K_3C_{60}$ measured for different pressures at a fixed temperature of T = 100 K.

All measured spectra have been consistently fitted using a Drude-Lorentz model, which included a free-carrier (*i.e.* Drude) term centered at $\omega = 0$ and a Lorentz oscillator at $\omega_{0,osc} \simeq$ 50 - 100 meV, which reproduced the mid-infrared absorption:

$$\sigma_1(\omega) + i\sigma_2(\omega) = \frac{\omega_p^2}{4\pi}\frac{1}{\gamma_D - i\omega} + \frac{\omega_{p,osc}^2}{4\pi}\frac{\omega}{i(\omega_{0,osc}^2 - \omega^2) + \gamma_{osc}\omega} \quad (1)$$

Here $\omega_p$ and $\gamma_D$ are the Drude plasma frequency and scattering rate, while $\omega_{p,osc}^2$, $\gamma_{osc}$, and $\omega_{0,osc}$ stand for the oscillator strength, damping coefficient and resonant frequency of the Lorentz term.

In Supplementary Table S1 we report all parameters extracted from those fits. For each given temperature, we observe a monotonous increase in the low-frequency extrapolation of the real part of the optical conductivity, $\sigma_0 = \omega_p^2/(4\pi\gamma_D)$, with increasing hydrostatic pressure. This is the expected behavior for a conducting material in which the electronic bandwidth gets enhanced by compressing the crystal lattice. No defined trend is observed instead for the high-frequency absorption at $\omega \simeq$ 50 - 100 meV.

The equilibrium data reported here have been used, for each given temperature and pressure, to normalise the transient optical spectra of $K_3C_{60}$ measured after photo-excitation, enabling the retrieval of all response functions of the perturbed material, as discussed in detail in the following sections.

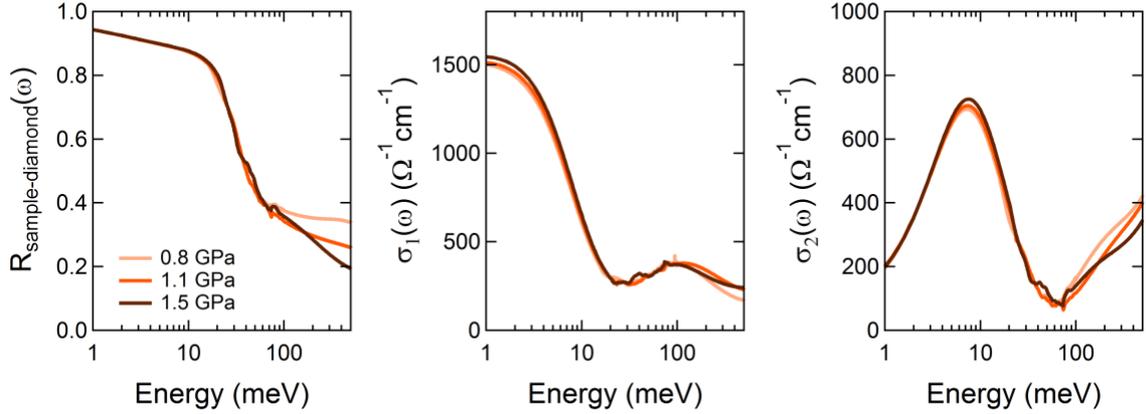

**Supplementary Figure S2. Equilibrium optical properties of $K_3C_{60}$ under pressure.** Equilibrium reflectivity at the sample-diamond interface (left), real (middle) and imaginary part (right) of the optical conductivity of $K_3C_{60}$, measured at T = 100 K for selected hydrostatic pressure values. Experimental data have been merged with extrapolations based on Drude-Lorentz fits below ~5 meV.

| T(K) | P(GPa) | $\sigma_0$ ($\Omega^{-1}cm^{-1}$) | $\omega_p$ (meV) | $\gamma_D$ (meV) | $\omega_{0,osc}$ (meV) | $\omega_{p,osc}$ (meV) | $\gamma_{osc}$ (meV) |
|---|---|---|---|---|---|---|---|
| 100 | 0.8 ± 0.1 | 1520 ± 20 | 300.2 ± 0.7 | 8.0 ± 0.1 | 76 ± 8 | 604.0 ± 0.6 | 160 ± 4 |
| 100 | 1.1 ± 0.1 | 1540 ± 20 | 307 ± 1 | 8.2 ± 0.1 | 80 ± 10 | 633 ± 1 | 190 ± 6 |
| 100 | 1.5 ± 0.1 | 1590 ± 30 | 313 ± 2 | 8.3 ± 0.1 | 70 ± 10 | 523 ± 2 | 165 ± 9 |
| 200 | 0.6 ± 0.1 | 1420 ± 20 | 295.0 ± 0.7 | 8.3 ± 0.1 | 75 ± 8 | 569.0 ± 0.7 | 154 ± 4 |
| 200 | 0.9 ± 0.1 | 1460 ± 30 | 308 ± 1 | 8.7 ± 0.1 | 81 ± 9 | 599.1 ± 0.9 | 155 ± 5 |
| 200 | 1.0 ± 0.1 | 1480 ± 20 | 300 ± 1 | 8.1 ± 0.1 | 67 ± 9 | 540 ± 1 | 150 ± 5 |
| 300 | 0.2 ± 0.1 | 1340 ± 20 | 290.8 ± 0.7 | 8.5 ± 0.1 | 70 ± 7 | 503.3 ± 0.6 | 141 ± 4 |
| 300 | 0.7 ± 0.1 | 1390 ± 40 | 300 ± 2 | 8.8 ± 0.2 | 86 ± 2 | 600 ± 20 | 200 ± 10 |
| 300 | 1.1 ± 0.1 | 1400 ± 20 | 298 ± 1 | 8.5 ± 0.1 | 66 ± 1 | 440 ± 9 | 135 ± 6 |

**Supplementary Table S1. Drude-Lorentz fit parameters of the equilibrium data.** Parameters extracted from Drude-Lorentz fits to the equilibrium optical response functions of $K_3C_{60}$ measured at different temperatures and pressures. The parameters relative to the Drude term (plasma frequency $\omega_p$, carrier scattering rate $\gamma_D$, and extracted zero-frequency conductivity $\sigma_0$) are displayed along with those of the mid-infrared absorption (center frequency $\omega_{0,osc}$, oscillator strength $\omega_{p,osc}$, and damping coefficient $\gamma_{osc}$). The pressure values were determined by fits of the ruby fluorescence line, as explained in Refs. 2,3.

**S3. Time resolved THz spectroscopy under pressure**

The transient optical response of $K_3C_{60}$ after photo-excitation was measured under hydrostatic external pressure by means of time-resolved THz spectroscopy. Hydrostatic pressure was applied using a membrane-driven Diacell®Bragg-LT diamond anvil cell (DAC) from Almax easyLab, designed for cryogenic experiments. This DAC allowed varying the static pressure by actuating a membrane with helium gas, without removing the cell from its holder and ensuring maximum reliability and reproducibility of the measurement conditions. The DAC was mounted on the cold finger of a liquid helium cryostat, where the sample temperature was accurately measured using a ruby thermometer[3].

We used diamond anvils of type IIac with Boehler-Almax design and a culet size of 2 mm. $K_3C_{60}$ powders (100 – 400 nm average grain size) were directly loaded in a 1.2 mm diameter sample compartment drilled in a 100 μm thick pre-indented Cu-gasket. Despite the relatively large sample compartment compared to the culet size of the anvils, no significant deformations were observed during the experiment. *In-situ* measurements of ruby R1 fluorescence line were used as a pressure gauge[2]. To avoid shifts in the pressure calibration axis due to different coupling of the fluorescence signal to the spectrometer, the fluorescence signal was passed through a single mode optical fiber. The overall accuracy of this measurement method is ~0.05 GPa. Pressure values as high as 2.5 GPa could be achieved. As in the case of the equilibrium characterisation (Supplementary Section S2), all sample handling operations were performed in an Ar atmosphere glove box with <0.1 ppm $O_2$ and $H_2O$ in order to avoid oxygen contamination.

$K_3C_{60}$ was excited using 300 fs long mid-infrared pulses with a spectrum centered at 7.3 μm wavelength (170 meV energy), tuned to resonance with an on-ball $T_{1u}$ vibration. These pump pulses were generated by difference frequency mixing of the signal and idler outputs of a two-stage optical parametric amplifier (OPA) in a 1 mm thick GaSe crystal. The OPA was pumped with 100 fs long pulses from a commercial Ti:Sapphire regenerative amplifier (800-nm wavelength). The pump pulses were then focused onto the sample, achieving a maximum fluence of ~3 mJ/cm$^2$, corresponding to peak electric fields of ~2.75 MV/cm.

The transient optical properties of K$_3$C$_{60}$ after photo-excitation were measured using THz *probe* pulses with spectral bandwidth extending from 1 to 7.1 THz, *i.e.* 4.1 - 29 meV.

The probe pulses were generated and detected in 200-µm thick <110> GaP, using 35-fs-long 800-nm pulses. The time resolution of the pump probe experiment is of about 300 fs, as determined by the bandwidth of the THz-TDS setup[7,8] and by the mid-infrared pump pulse duration.

Note that, given the geometrical constraints of the experimental setup, only the frequency components of the probe spectrum above ~1.6 THz (6.5 meV) could be focused on a spot smaller than 1.2 mm, corresponding to the size of the compressed K$_3$C$_{60}$ pellet. Therefore, unlike in the experiment of Ref. 6 (performed at ambient pressure and with a larger sample), here all frequencies smaller than the ~1.6 THz cut-off (6.5 meV) had to be excluded from our analysis, being most likely affected by spurious reflections from the highly reflective Cu-gasket surrounding the sample.

## S4. Measurement of the THz pump-induced changes under pressure

The stationary electric field reflected by the sample, $E_R(t)$, and the pump-induced changes, $\Delta E_R(t,\tau) = E_R^{pumped}(t,\tau) - E_R(t)$, were simultaneously acquired at each time delay τ by filtering the electro-optic sampling signals with two lock-in amplifiers[9]. $\Delta E_R(t,\tau)$ and $E_R(t)$ were then independently Fourier transformed to obtain the complex-valued, frequency dependent $\Delta \tilde{E}_R(\omega, \tau)$ and $\tilde{E}_R(\omega)$.

The advantage of a double lock-in amplifier detection scheme relies on the simultaneous measurement of the light-induced changes and the reference electric field. This technique avoids the introduction of possible phase artifacts (e.g. due to long term drifts) and is particularly useful when the measured electric field contains fast-varying frequencies.

The raw pump-induced changes in reflected field amplitude, $|\Delta \tilde{E}_R(\omega, \tau)/\tilde{E}_R(\omega)|$, measured at T = 100 K and for a time delay $\tau = 1$ ps, are reported in Supplementary Fig. S3 for different applied pressures. Full spectra are shown on the left, while the pressure dependence of frequency-integrated values is displayed on the right.

Remarkably, the same trend found in the $\sigma_1(\omega)$ spectral weight loss in main text Fig. 4 (*i.e.* a reduction of the light-induced effect with increasing pressure), is also evident in the raw, unprocessed reflectivity changes reported here. This suggests that the main conclusion of our experiment, a reduction in the transient gap size with pressure, is clearly seen independent of any processing protocol.

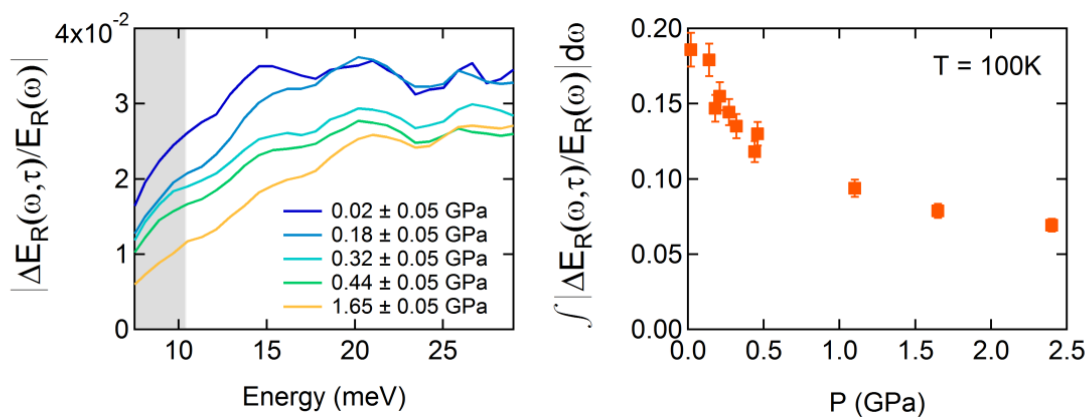

**Supplementary Figure S3. Raw data analysis of $K_3C_{60}$ at 100K**. (**left panel**) Evolution of the raw $|\frac{\Delta \tilde{E}_R(\omega, \tau)}{\tilde{E}_R(\omega)}|$ for selected values of external pressure, measured in $K_3C_{60}$ at 100K, at $\tau = 1$ ps after excitation. (**right panel**) Pressure dependence of the spectrally-integrated values. In analogy with main text Fig. 4, the integration was performed in the 6.5 – 12.9 meV range (gray shaded area in left panel).

## S5. Determination of the transient optical properties.

The complex reflection coefficient of the photo-excited material, $\tilde{r}(\omega, \tau)$, was determined using the relation:

$$\frac{\tilde{r}(\omega, \tau) - \tilde{r}_0(\omega)}{\tilde{r}_0(\omega)} = \frac{\Delta \tilde{E}_R(\omega, \tau)}{\tilde{E}_R(\omega)} \qquad (2)$$

To calculate these ratios, the static reflection coefficient $\tilde{r}_0(\omega)$ was extracted at all temperatures and pressures from the equilibrium optical properties (see Supplementary Section S2).

These "raw" light-induced changes were then reprocessed to take into account the penetration depth mismatch between THz probe (700 - 800 nm) and mid-infrared pump (170 - 220 nm). Importantly, this renormalization only affects the size of the response, whereas the qualitative changes in the optical properties are independent of it and of the specific model chosen to account for penetration depth mismatch. A precise reconstruction method consists in treating the excited surface as a stack of thin layers with a homogeneous refractive index and describing the excitation profile by an exponential decay (see Supplementary Fig. S4). By calculating the coupled Fresnel equations of such multi-layer system, the complex refractive index at the surface, $\tilde{n}(\omega, \tau)$, can be retrieved, and from this the complex conductivity for a volume that is homogeneously transformed,

$$\tilde{\sigma}(\omega, \tau) = \frac{\omega}{4\pi i}[\tilde{n}(\omega, \tau)^2 - \varepsilon_\infty]. \qquad (3)$$

Note that, in order to minimize the effects of pump-probe time resolution due to a finite duration of the probe pulse, we operate the delay stages in the setup as explained in Ref. 7. Therefore, as already mentioned in Supplementary Section S3, our temporal resolution is limited only by the duration of the pump pulse and by the inverse bandwidth of the probe pulse. As discussed already in Ref. 6, for all measurements presented here the time resolution is of the order of 300 fs (see also Supplementary Section S3), the mid-infrared pump pulses are also ~300 fs long, transient coherence is induced in 1 ps, and the relaxation occurs within 2 – 10 ps. Therefore all possible spectral deformations are negligible[10,11].

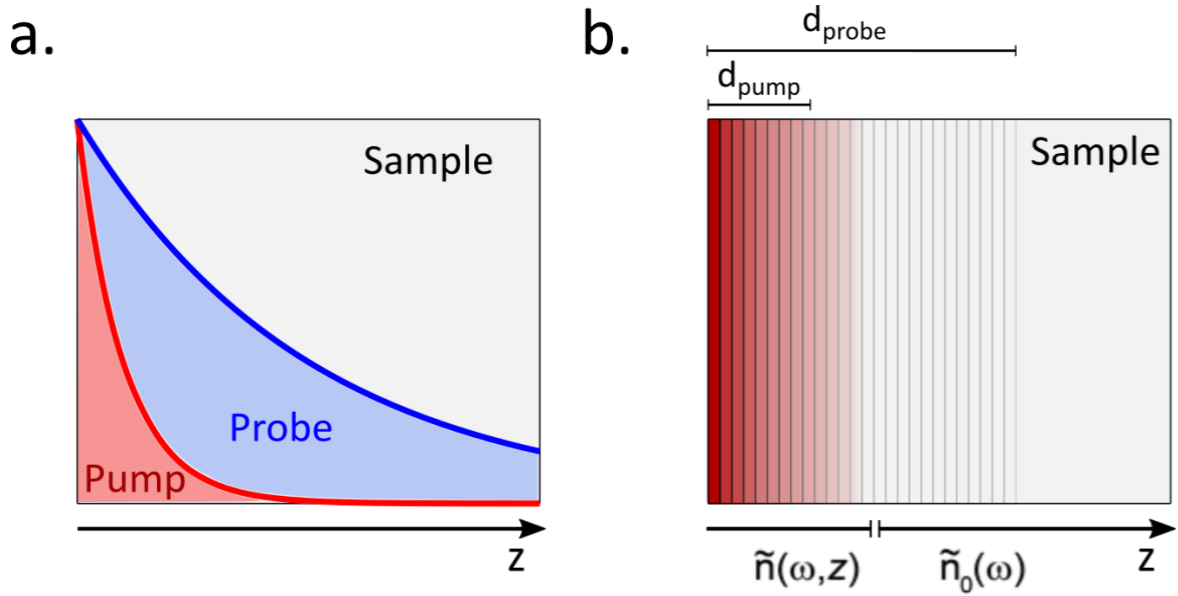

**Supplementary Figure S4. Model for penetration depth mismatch**. (**a**) Schematics of pump-probe penetration depth mismatch. (**b**) Multi-layer model with exponential decay used to calculate the pump-induced changes in the complex refractive index $\tilde{n}(\omega)$. The transition from red to background (gray) represents the decaying pump-induced changes in $\tilde{n}(\omega)$.

**S6. Influence of the choice of the penetration-depth mismatch model**

As explained in Supplementary Section S5, the transient optical response functions reported in the main text have been reconstructed through a multi-layer model, in order to account for the pump-probe penetration depth mismatch. This model treats the excited sample as a stack of thin layers with a homogeneous refractive index, assuming an exponential excitation profile (see Supplementary Fig. S4). A valid, but less accurate, alternative to such model can be obtained by considering the excited material as a single, uniform layer of thickness equal to the pump penetration depth, on top of an unperturbed bulk. As a benchmark, we extracted the transient optical properties using these two different models, starting from the "raw" light-induced changes measured at 1 ps pump–probe delay, T = 100 K, P = 0 GPa. Supplementary Fig. S5 shows the results of this comparison. The transient optical properties retrieved in the single-layer approximation (dark blue curves), and those obtained using a multi-layer model (light blue curves) only differ slightly. The qualitative features in the optical properties, e.g. the presence of a light-induced gap in the real part of the optical conductivity, are clearly independent on the model chosen to account for the penetration depth mismatch.

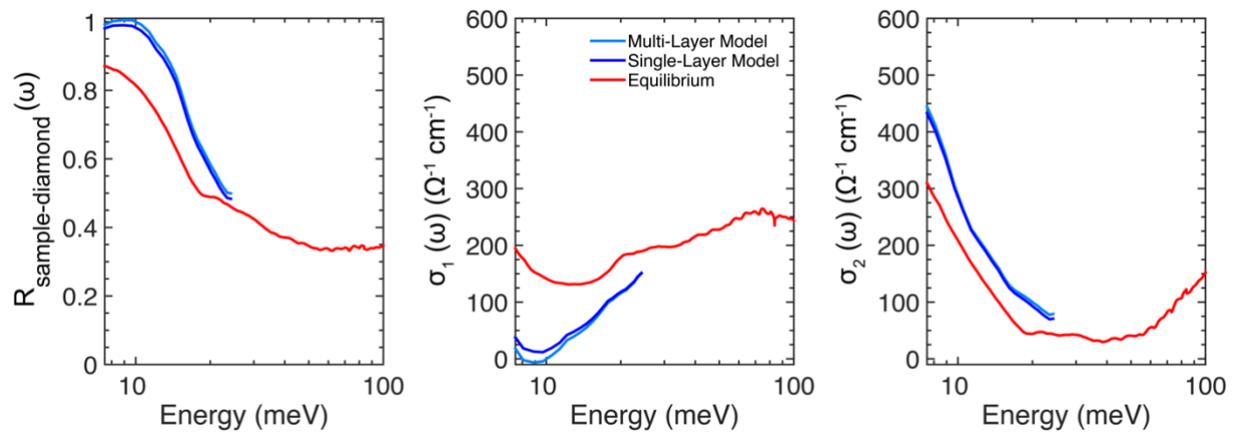

**Supplementary Figure S5. Multi-layer model vs Single-layer model.** Reflectivity, R(ω), and complex optical conductivity, σ(ω), of $K_3C_{60}$ measured at τ = 1 ps pump–probe delay, T = 100 K, P = 0 GPa. Light-blue curves are obtained using the single-layer model; dark-blue curves are retrieved using the multi-layer model.

### S7. Influence of uncertainties in the equilibrium optical properties

The error on the reconstructed transient optical response of photo-excited $K_3C_{60}$ are primarily determined by (i) the uncertainty in the absolute value of the measured equilibrium reflectivity, which is typically of the order of ±1%, and (ii) by the value of the pump penetration depth $d_{pump}$ used to reconstruct the $\tilde{n}(\omega, z)$ profile in the multilayer model. Similar to Supplementary Section S6, we use as benchmark the transient optical properties reconstructed from the recorded "raw" light-induced changes measured at 1 ps pump–probe delay, T = 100 K, P = 0 GPa.

Supplementary Fig. S6a and S6b show as colored bands, the propagated error bars in the transient optical properties for maximum uncertainties of ±2% in the equilibrium R(ω) and of ±10% in $d_{pump}$, respectively. Importantly, all qualitative features in the retrieved response of the perturbed material, as for example the presence of a light-induced gap in the real part of the optical conductivity, remain unaffected by those uncertainties in the equilibrium optical properties.

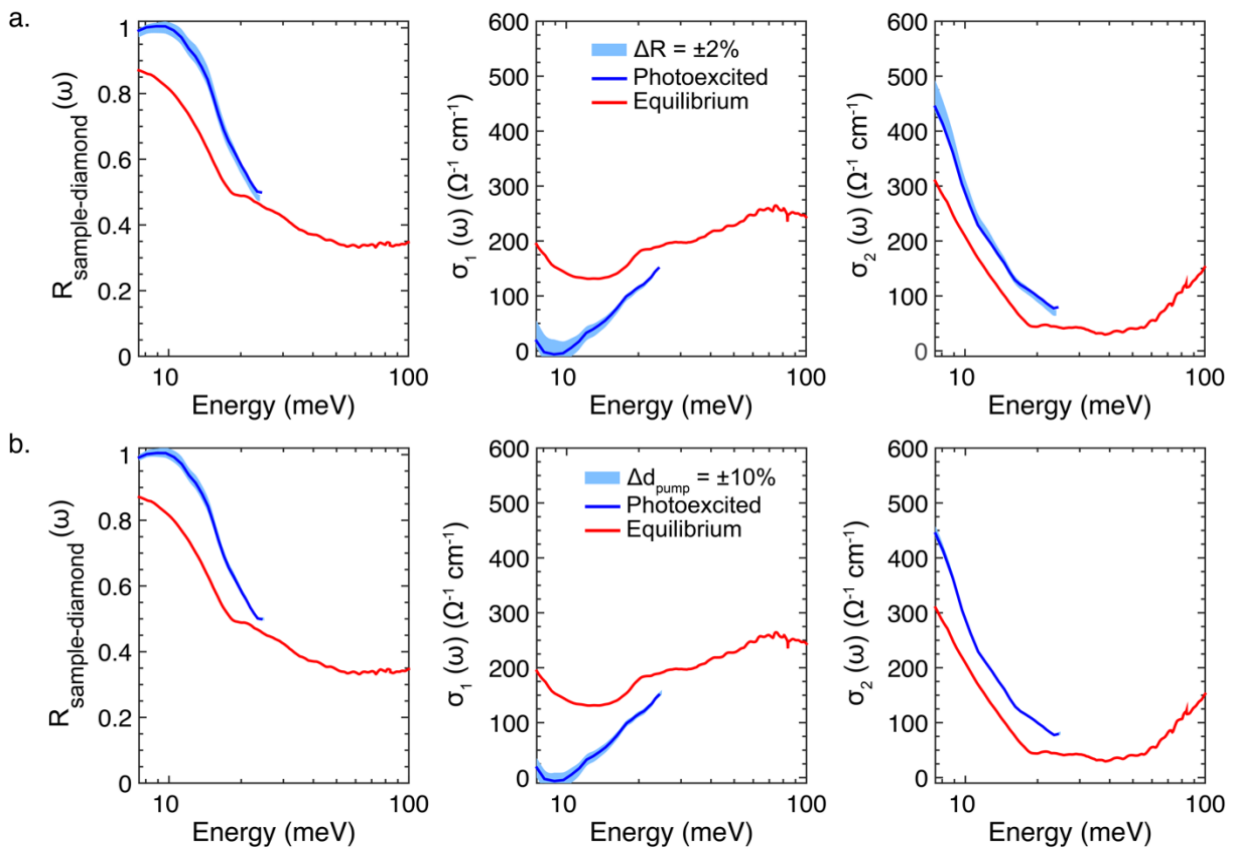

**Supplementary Figure S6. Error in the retrieved transient optical properties originated by uncertainties in the equilibrium optical properties.** The dark-blue curves are the reflectivity and complex optical conductivity of K$_3$C$_{60}$ measured at τ = 1 ps pump–probe delay, T = 100 K, P = 0 GPa. Error bars, shown as colored bands, have been propagated from **(a)** a ±2% uncertainty in the value of the equilibrium R(ω), and **(b)** from a ±10% change in the pump penetration depth.

## S8. Time evolution of the transient optical response of $K_3C_{60}$ under pressure

The temporal evolution of the light-induced state in $K_3C_{60}$ does not show a strong pressure dependence and resembles that observed under ambient conditions[6]. Supplementary Fig. S7 shows the frequency averaged "raw" reflectivity changes as function of pump-probe time delay for selected values of external pressure. Those curves have been obtained by sampling the peak of the THz-probe electric field at different time delays before and after excitation. Independent of the specific value of the external pressure, the light-induced reflectivity changes are always positive (see also Fig. 3 of the main text). All curves in Supplementary Fig. S7 have been fitted with a double exponential decay with relaxation time constants $\tau_1 \sim 1$ ps and $\tau_2 \gg 10$ ps.

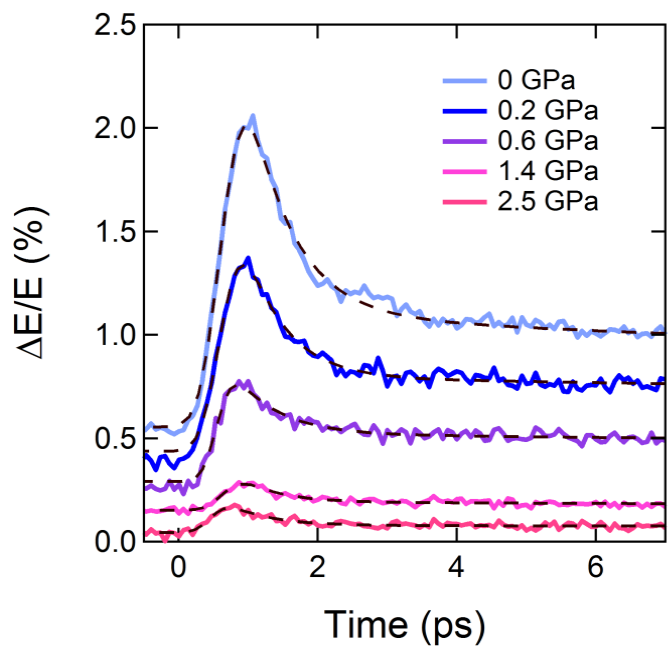

**Supplementary Figure S7. Time evolution of the photo-excited state in $K_3C_{60}$ under pressure.** Frequency averaged changes in the reflected THz field as function of time delay, for selected values of external pressure, measured at T = 100 K. The experimental data (solid lines) have been obtained by sampling the peak of the THz-probe electric field as a function of pump-probe delay. Dashed lines are exponential fits. Rigid vertical offsets have been introduced for clarity.

Supplementary Fig. S8 shows the transient optical properties of $K_3C_{60}$ measured at T = 300 K and P = 0.03 GPa for four representative pump-probe delays. The light-induced changes in the optical properties are largest at early times ($\tau$ = 0.8, 1 ps) and become progressively smaller at later delays ($\tau$ = 2 ps). Similar data have already been reported for $K_3C_{60}$ at ambient pressure in Ref. 6.

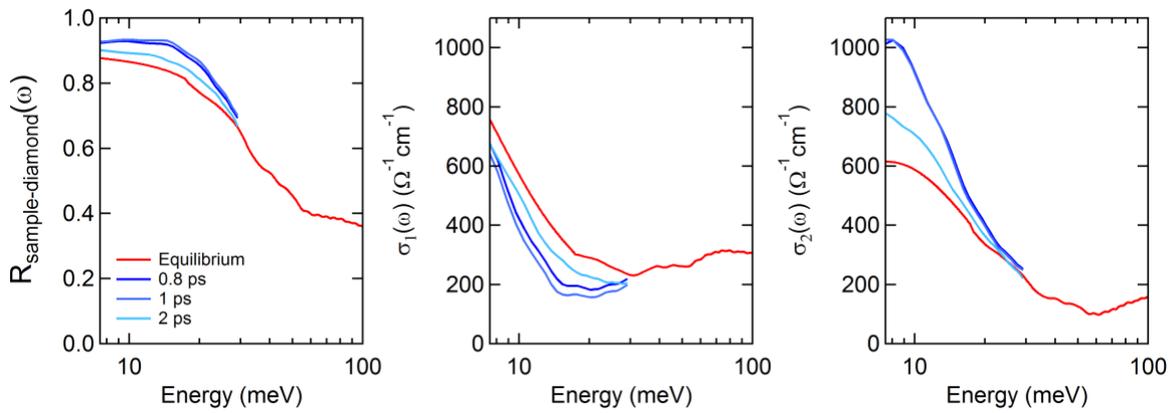

**Supplementary Figure S8. Transient optical properties of $K_3C_{60}$ at different pump-probe delays.** Reflectivity at the sample-diamond interface (left), real (middle) and imaginary part (right) of the optical conductivity of $K_3C_{60}$, measured at T = 300 K and P = 0.03 ± 0.05 GPa for selected pump-probe delays. All data were taken with the same pump fluence (3 mJ/cm$^2$).

**S9. Pressure-dependent transient optical properties at T = 200 K and T = 300 K.**

The pressure dependence of the transient optical properties of $K_3C_{60}$, measured at $\tau$ = 1 ps and T = 100 K, is displayed in Fig. 3 of the main text. Supplementary Fig. S9 and S10 show additional data set measured at higher temperatures, T = 200 K and T = 300 K, respectively. For both temperatures, the ambient pressure response shows a partial light-induced gapping in $\sigma_1(\omega)$, accompanied by a low-frequency divergence in $\sigma_2(\omega)$. On the other hand, at higher pressures, the photo-induced changes are gradually suppressed and the measured transient conductivity evolves into that of a metal with enhanced carrier mobility (narrower Drude peak).

As discussed in the main text, in this high temperature regime, a high-mobility metallic state was proposed in Ref. 6 to interpret the ambient pressure data. However, this interpretation was also not unique, as a superconducting-like state with progressively lower coherence would have also been consistent with the measured spectra. Figure 5b-c of the main text, whose data points have been extracted from fits to the data of Supplementary Fig. S9 and S10, clearly shows that in the low-pressure regime a decreasing zero-frequency conductivity, $\sigma_0$, with increasing pressure is retained all the way to T = 300 K, suggesting that some incipient features of a transient superconducting state may already be present up to room temperature.

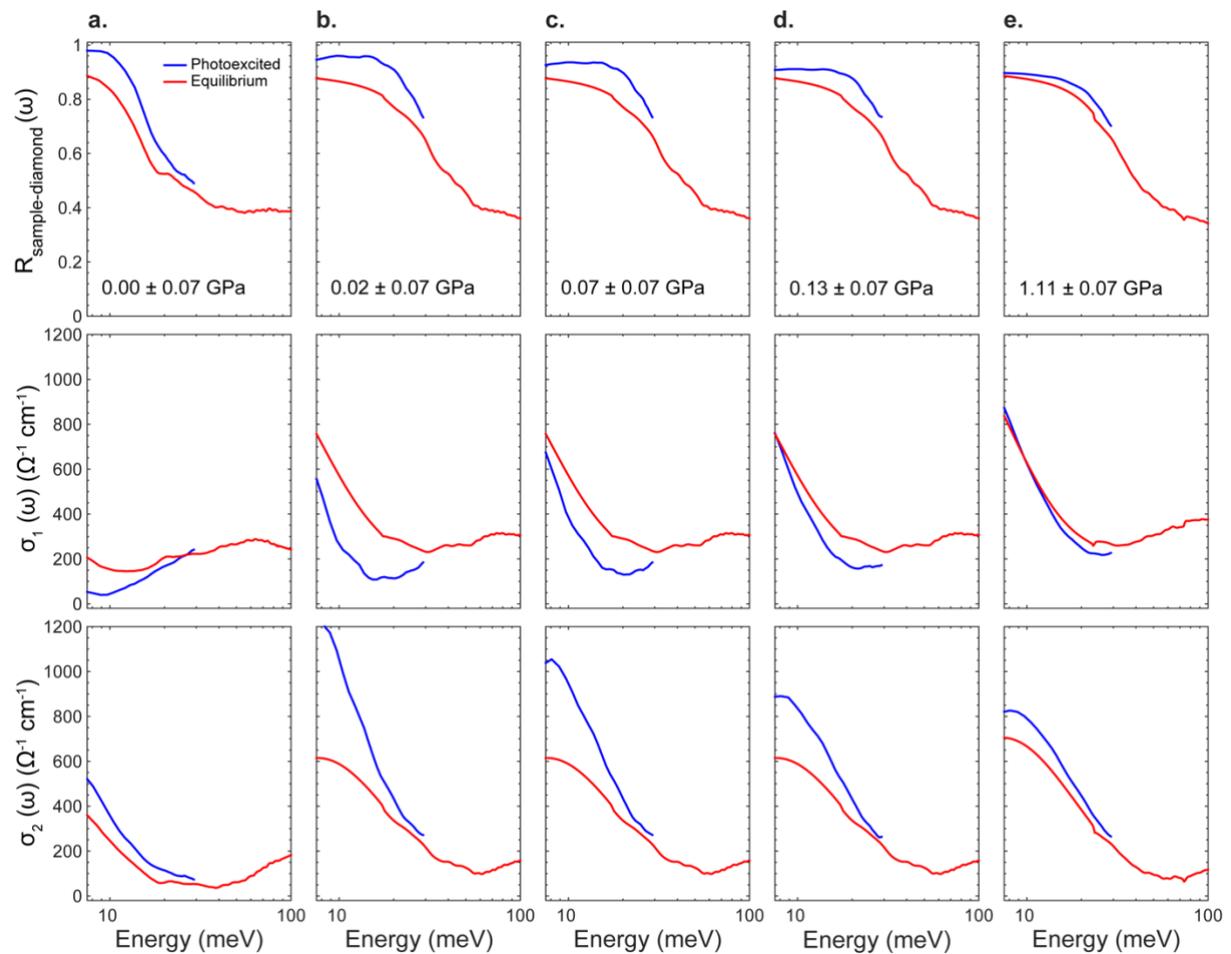

**Supplementary Figure S9. Pressure dependence of the transient optical properties of $K_3C_{60}$ at T = 200 K.** Reflectivity at the sample-diamond interface and complex optical conductivity of $K_3C_{60}$ measured at equilibrium (red) and 1 ps after photoexcitation (blue) at T = 200 K, for different external hydrostatic pressures. All data were taken with the same pump fluence (3 mJ/cm$^2$).

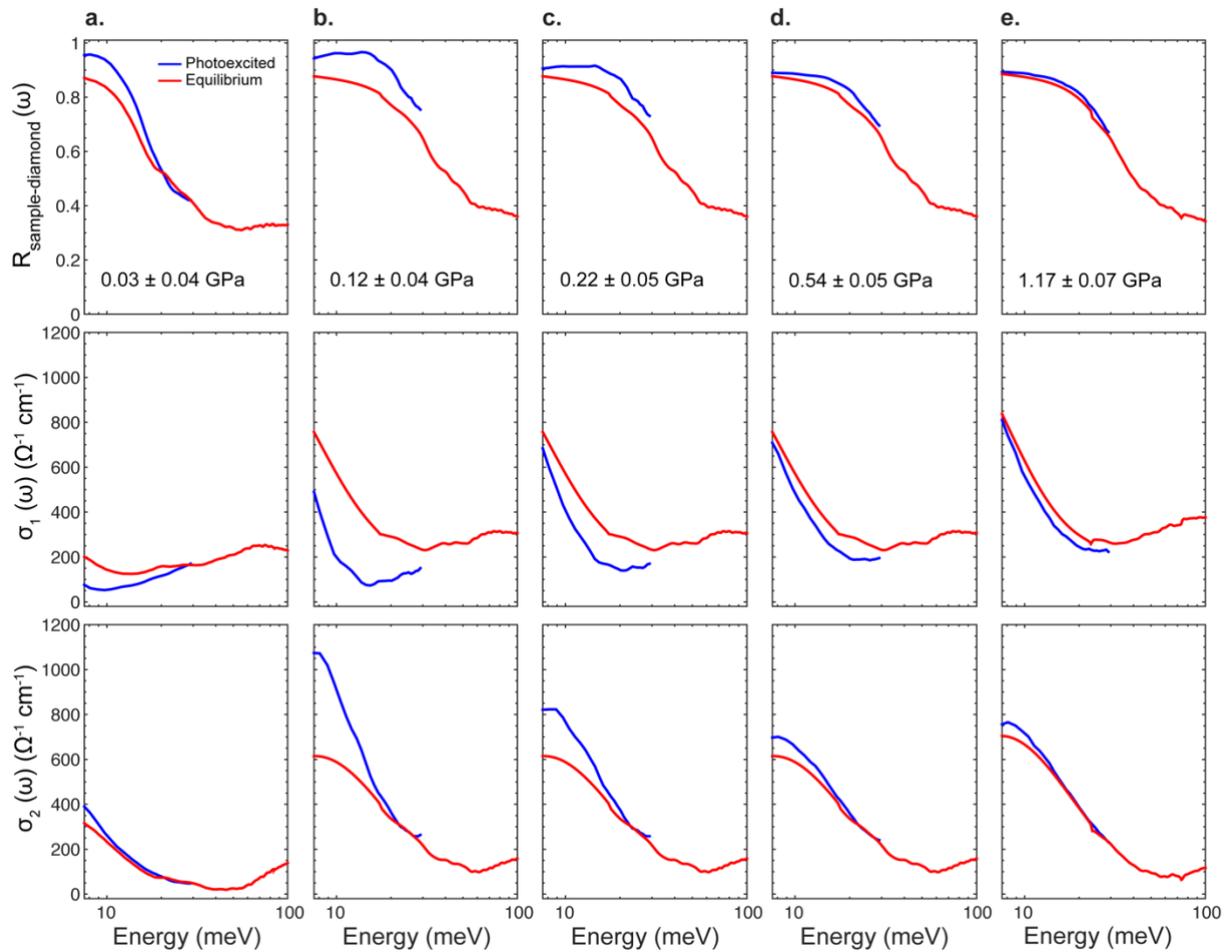

**Supplementary Figure S10. Pressure dependence of the transient optical properties of $K_3C_{60}$ at T = 300 K.** Reflectivity at the sample-diamond interface and complex optical conductivity of $K_3C_{60}$ measured at equilibrium (red) and 1 ps after photoexcitation (blue) at T = 300 K, for different external hydrostatic pressures. All data were taken with the same pump fluence (3mJ/cm$^2$).

**S10. Drude-Lorentz fits of the out-of-equilibrium optical response**

The transient optical properties of $K_3C_{60}$ were modeled at all measured temperatures and pressures by fitting the perturbed reflectivity and complex optical conductivity simultaneously, using the same Drude-Lorentz model which was employed to fit the equilibrium response (see Supplementary Section S2):

$$\sigma_1(\omega) + i\sigma_2(\omega) = \frac{\omega_p^2}{4\pi}\frac{1}{\gamma_D - i\omega} + \frac{\omega_{p,osc}^2}{4\pi}\frac{\omega}{i(\omega_{0,osc}^2 - \omega^2) + \gamma_{osc}\omega} \quad (4)$$

This model is obviously expected to fully capture the metallic character of the transient optical properties measured at high pressure. In addition, while more accurate modelling has been used in Ref. 6 to describe the transient superconducting-like response of K$_3$C$_{60}$, it is worth noting that even the simple $\gamma_D \to 0$ limit of the Drude conductivity used here can capture the response of a superconductor below gap:

$$\sigma_1(\omega) + i\sigma_2(\omega) = \frac{\pi}{2}\frac{N_S e^2}{m}\delta[\omega = 0] + i\frac{N_S e^2}{m}\frac{1}{\omega}. \quad (5)$$

Here $N_S$, $e$, and $m$ are the superfluid density, electron charge, and electron mass, respectively. Importantly, the transient nature of a superconductor with finite lifetime appears in the above formula as a broadening in the linewidth of the zero-frequency Dirac delta[12].

In Supplementary Fig. S11 we report two representative fits to 100 K data measured at ambient pressure and at 0.26 ± 0.05 GPa, which all show a good agreement with the experiment. Both transient data sets (blue curves) could be fitted with the Drude model described in Eq. 4 using as initial guess the fit parameters of the corresponding equilibrium data sets (red curves). Fits were performed by either keeping the high-frequency (> 40 meV) mid-infrared oscillator fixed (grey dashed line), or by letting it evolve freely (black dash-dotted line). Remarkably, both fits returned very similar results, with identical Drude parameters within the error bars. We therefore chose to rely on the model with the least number of free parameters (gray dashed lines).

In Supplementary Table S2 we show the Drude parameters for all data sets (extracted from the fits with fixed mid-infrared absorption band). Note that the calculated zero-frequency extrapolations of the optical conductivity, $\sigma_0$, reported in this table are also plotted and discussed in the main text (see Fig. 5).

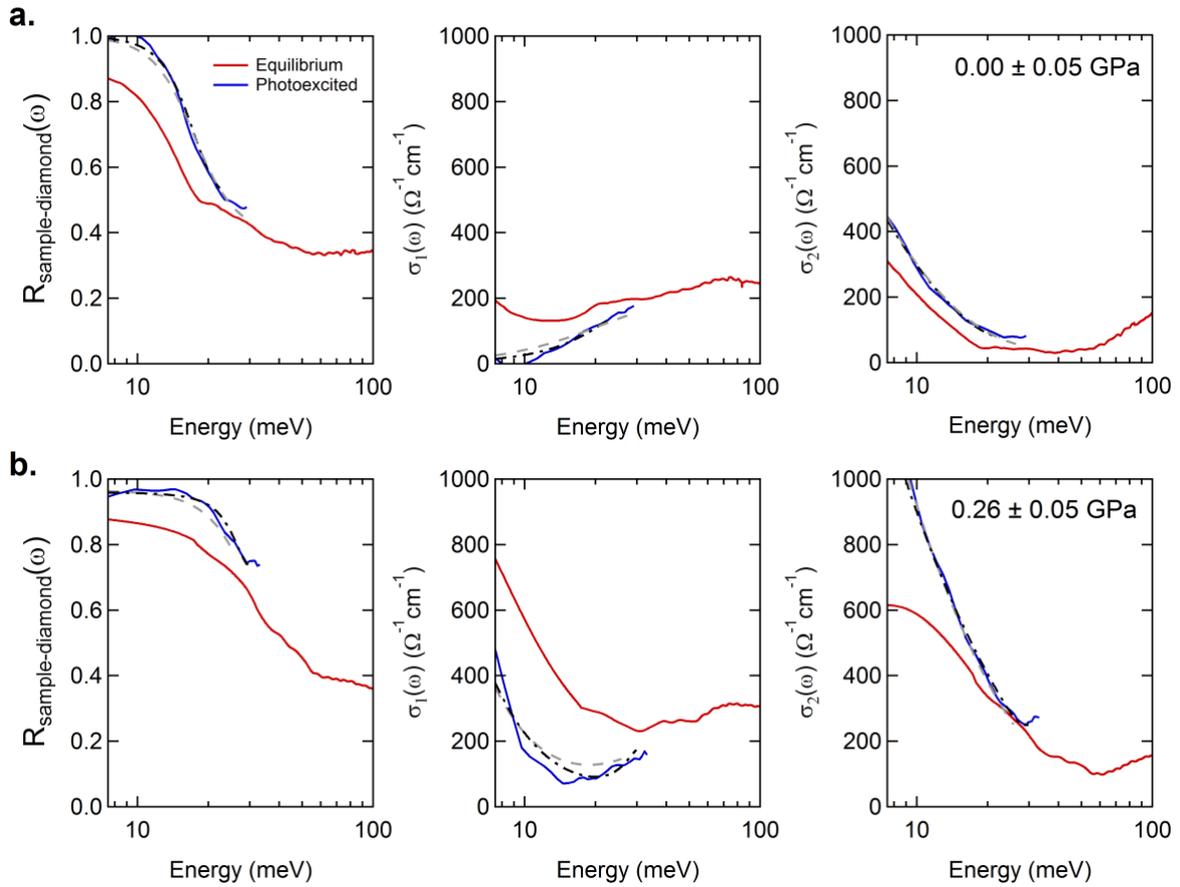

**Supplementary Figure S11. Drude-Lorentz fits to the transient optical response.** Reflectivity at the sample-diamond interface and complex optical conductivity of $K_3C_{60}$ measured at equilibrium (red) and 1 ps after photoexcitation (blue) for T = 100 K at ambient pressure (**a**) and at P = 0.26 ± 0.05 GPa (**b**). Experimental data (full lines) are displayed along with Drude-Lorentz fits, performed by either keeping the high-frequency (> 40 meV) mid-infrared oscillator fixed (grey dashed line), or by letting it evolve freely (black dash-dotted line).

| T(K) | P(GPa) | $\sigma_0$ ($\Omega^{-1}$cm$^{-1}$) | $\omega_P$ (meV) | $\gamma_D$ (meV) |
|---|---|---|---|---|
| 100 | 0.02 ± 0.05 | $10^{17} \pm 10^{13}$ | 169 ± 3 | 0.0 ± 0.3 |
| 100 | 0.05 ± 0.05 | $10^{18} \pm 10^{12}$ | 280 ± 4 | 0.0 ± 0.3 |
| 100 | 0.14 ± 0.05 | 17000 ± 8000 | 279 ± 4 | 0.6 ± 0.3 |
| 100 | 0.18 ± 0.05 | 7000 ± 1000 | 276 ± 3 | 1.4 ± 0.3 |
| 100 | 0.21 ± 0.05 | 8000 ± 1000 | 276 ± 3 | 1.3 ± 0.2 |
| 100 | 0.27 ± 0.05 | 6200 ± 800 | 277 ± 3 | 1.7 ± 0.2 |
| 100 | 0.32 ± 0.05 | 5100 ± 500 | 278 ± 3 | 2.0 ± 0.2 |
| 100 | 0.44 ± 0.05 | 4800 ± 600 | 263 ±3 | 2.0 ± 0.3 |
| 100 | 0.46 ± 0.05 | 3900 ± 300 | 281 ±6 | 2.7 ± 0.2 |
| 100 | 1.10 ± 0.05 | 3900 ± 400 | 284 ± 3 | 2.8 ± 0.3 |
| 100 | 1.70 ± 0.06 | 3100 ± 300 | 294 ± 4 | 3.7 ± 0.3 |
| 100 | 2.40 ± 0.09 | 3700 ± 300 | 310 ± 4 | 3.5 ± 0.3 |
| 200 | 0.00 ± 0.07 | $10^{18} \pm 10^{12}$ | 191 ± 1 | 0.0 ± 0.1 |
| 200 | 0.02 ± 0.06 | 4500 ± 300 | 293 ± 2 | 2.5 ± 0.2 |
| 200 | 0.07 ± 0.07 | 3000 ± 200 | 293 ± 2 | 3.8 ± 0.2 |
| 200 | 0.13 ± 0.07 | 2100 ± 100 | 293 ± 2 | 5.1 ± 0.2 |
| 200 | 0.22 ± 0.07 | 2020 ± 80 | 295 ± 2 | 5.8 ± 0.2 |
| 200 | 0.41 ± 0.07 | 1790 ± 50 | 298 ± 2 | 6.7 ± 0.2 |
| 200 | 1.10 ± 0.07 | 1920 ± 40 | 310 ± 2 | 6.7 ± 0.1 |
| 300 | 0.03 ± 0.04 | 4700 ± 800 | 169 ± 1 | 0.8 ± 0.1 |
| 300 | 0.03 ± 0.04 | 2700 ± 100 | 280 ± 2 | 3.8 ± 0.2 |
| 300 | 0.07 ± 0.04 | 4900 ± 400 | 275 ± 2 | 2.0 ± 0.2 |
| 300 | 0.12 ± 0.04 | 4500 ± 400 | 275 ± 3 | 2.3 ± 0.2 |
| 300 | 0.23 ± 0.05 | 2200 ± 100 | 277 ± 3 | 4.7 ± 0.2 |
| 300 | 0.54 ± 0.05 | 1660 ± 50 | 279 ± 2 | 6.3 ± 0.2 |
| 300 | 1.17 ± 0.07 | 1750 ± 30 | 297 ± 1 | 6.8 ± 0.1 |

**Supplementary Table S2. Drude parameters extracted from fits to the transient data.** Parameters extracted from Drude-Lorentz fits to the transient optical response functions of $K_3C_{60}$ measured at different temperatures and pressures. The plasma frequency $\omega_P$, Drude scattering rate $\gamma_D$, and extracted dc conductivity $\sigma_0$ are displayed. The pressure values are extracted by fits of the ruby fluorescence line, as explained in Refs. 2,3.

# REFERENCES (Supplementary Information)